\documentstyle[prb,preprint,aps]{revtex}
\begin{document}
\draft
\date{July 12, 1995}
\title
{Spin fluctuations of a random vacuum state:\\
a new approach to the Kondo lattice}
\author
{J.M. Prats and F. L\'opez-Aguilar}
\address
{Departament de F\'\i sica, Grup d'Electromagnetisme,
Universitat Aut\`onoma de Barcelona,\\
Bellaterra, E-08193 Barcelona, Spain}
\maketitle
\begin{abstract}
We address the Kondo lattice with one conduction band by introducing
a representation of the ($s=1/2$) local moments consisting in an
$s=1$ triplet of spin fluctuations of a vacuum state. The excitation
of these fluctuations by the electronic
field results in the creation
of new fermionic modes. It is shown that, in the one-electron case,
the zero-temperature spectrum of these excitations
consists in two $s=1/2$ doublet bands and
an $s=3/2$ quadruplet. One of the doublets is seen to be
a slight modification of the bare initial electronic band for low
values of the exchange coupling,
while the other bands correspond to
complex, coherent recombinations of the bare electron and spin
fluctuations displaying large effective masses.
\end{abstract}


\section {Introduction}

Heavy-fermion metals display a rich phenomenology produced by the
interaction between conduction electrons and partially filled
$f$-bands from (typically) Ce or U.\cite{fisk,stewart} Its
high-temperature properties are those of a collection of weakly
interacting local moments and conduction electrons with quite
ordinary masses. At low temperatures, however, they exhibit Fermi
liquid behaviour with values of the linear specific heat and
temperature-independent Pauli spin susceptibility corresponding to
huge quasiparticle masses. This Fermi liquid becomes, in many
systems, antiferromagnetic, superconducting, or both.

The high-temperature behaviour described above, suggests that these
systems should be represented by the
{\em Kondo lattice model\/} (KLM),
in which the conduction electrons are coupled to the local moments
through an exchange interaction. This model is the limit of the more
general {\em Anderson lattice model\/} (ALM)
when the fluctuations in
the number of $f$-electrons per Ce or U site can be neglected.
\cite{schrieffer}

A number of techniques have been proposed for dealing with these
models (especially the ALM).\cite{lee,fulde} Of particular
importance have been the mean-field \cite{coleman} and variational
\cite{rice} approximations. In both cases, the Hamiltonian of the
ALM is essentially transformed into an effective Hamiltonian
in which the $f$-electrons are uncorrelated, shifted close to the
Fermi level and hybridize weakly with the conduction electrons.
This high density of states at the Fermi level would explain the
high values of the linear coefficient of the specific heat and the
magnetic susceptibility. The formulation of these theories relies
heavily on the fact that the occupation of the $f$-orbitals is not
exactly integer.

In this article, we present a new approach to the description of the
heavy-fermion state. By selecting the KLM, we are effectively
asserting that the phenomenology of these systems is associated
with the interaction of conduction electrons and local moments
and that the $f$-charge fluctuations do not play an essential role.
The general philosophy of our approach bears strong parallelisms
with the physical discussion of the Kondo model, namely, the
interaction of a single magnetic impurity with conduction electrons
through an antiferromagnetic exchange coupling.\cite{mahan}
In this model, the excess magnetic susceptibility is seen to depart,
at a characteristic ($T_{k}$) temperature, from the Curie-Weiss
law, smoothly approaching a constant limit as $T \to 0$.
This is because at $T=0$, the local spin forms a collective
singlet state with the spins of the conduction band with a binding
energy of essentially $K_B T_k$.\cite{yosida}
Thus, for temperatures below $T_k$, the resonant
scattering of the conduction electrons acts
as an effective temperature for the local moment giving rise to
the mentioned $T$-dependence of the magnetic susceptibility.

Analogously, in the treatment of the KLM presented in this paper,
heavy fermions appear as collective, many-particle states made up
of electrons and spin fluctuations. Nevertheless, while the Kondo
resonance is an $s=0$ singlet, our collective excitations are
fermions. Another difference is that, while the Kondo resonance
can form only for antiferromagnetic couplings, the situation in
the KLM is the same for both ferro and antiferromagnetic couplings.
This is important since heavy-fermion metals like CeAl${}_3$ are
known to have ferromagnetic couplings.\cite{andres}

The key idea of this article is the introduction of a new, very
physical representation of the algebra of the local moments. Let
us suppose for a moment that the conduction electrons are absent.
If we consider a lattice of $N$ sites with a local $s=1/2$ spin
at each site, the Hilbert space of the system is spanned by a
basis of $2^N$ states corresponding to all the possible spin
arrangements constructed from the up and down states of the local
spin at each lattice site. Since each of these $2^N$ states has
the same energy when conduction electrons are absent, the global
state of the system can be regarded as a linear combination of
the $2^N$ basic vectors, each of them carrying the same probability
and an arbitrary, random phase. This is what we call the {\em
random vacuum state\/} (RVS). We show that by repeatedly applying
the spin operators $\sqrt{2} S^{+}_{fi}$, $\sqrt{2} S^{-}_{fi}$,
and $2 S^{z}_{fi}$ to this vacuum state we obtain, in the
thermodynamic limit, a set of orthonormal vectors that span a new
representation of the algebra of the local moments. The states
of this new Hilbert space can be interpreted as spin fluctuations
of the RVS. This new representation, initially constructed using
physical considerations, can be presented at the end in an
axiomatic fashion as a perfectly consistent, rigorous mathematical
structure.

Thus, with this representation, when conduction electrons are
incorporated into the system, their interaction with the local
moments can be viewed as an electronic excitation of spin
fluctuations of a vacuum state. In this process, the conduction
electrons get dressed with spin fluctuations giving rise to new
fermionic excitations. For low relative values of the exchange
coupling (the actual physical situation), we find two types of
excitations. One type corresponds to slight modifications of the
initial bare conduction electrons. The other modes are complex,
many-particle combinations of electrons and spin fluctuations
displaying large effective masses.

We provide a systematic technique to calculate the energy and
wave function of these excitations. The method essentially
consists in determining variationally the best approximation
to the actual wave function within the space in which the
bare electron dresses with $n$ spin fluctuations at most. The
larger we take $n$, the better the approximation.

In this basic, qualitative article, we consider the lowest
($n=1$) approximation for the system with a single conduction
electron only. Although the lowest approximation proves
to be inadequate for the precise determination of the
heavy-fermion bands, it does reveal the qualitative picture
presented in this paper. The precise calculation of the
heavy-fermion modes for low relative values of the exchange
coupling will require to consider higher orders ($n>1$) of
approximation and promises to be a hard technical task.

In Section II, we present the detailed construction of the
algebra of spin fluctuations. The determination of the
translational and spin rotational symmetries of the system
within this new representation is carried out in Section III.
The use of these symmetries will be essential for the
construction and characterization of the fermionic excitations.

In Section IV we address the simple case of a single conduction
electron interacting with the lattice of local moments and study
how this bare electron gets dressed with spin fluctuations. The
inspection of the atomic limit (band width $\to$ 0) leads to the
conclusion that the spectrum of the system contains two $s=1/2$
doublet bands and an $s=3/2$ quadruplet. These bands are then
calculated in the lowest approximation to determine their
qualitative features away from the atomic limit. As we already
mentioned, this approximation indicates that for small values
of the coupling, one of the doublets is a minor modification of
the bare conduction band, while in the other doublet and in the
quadruplet, the conduction electron strongly combines with spin
fluctuations forming bands with large effective masses. We also
find in this analysis, that the two doublets hybridize
{\em only\/} above a certain critical value of the coupling.

The natural extension of this picture to the case of a system
with a finite electronic density would require to treat the bare
conduction electrons as a Fermi sea with electrons and holes as
elementary excitations with positive energies. These electrons
and holes will be the ones giving rise to collective modes when
they combine with spin fluctuations. The existence of collective
states with negative energies would lead to the condensation of
some of these modes in the ground state,
forming the heavy-fermion
gas that is observed experimentally. The detailed study of the
Kondo lattice with a finite density of conduction electrons
within the framework of the approach introduced in this article
is intended to be the focus of our future endeavours.

\section {Spin fluctuations of a random vacuum state}

We shall study a KLM consisting in a conduction band coupled to
a lattice of $s=1/2$ local momemts by an exchange interaction.
The Hamiltonian is given by
\begin{equation}
H=\varepsilon_{{\bf k}} c^{\dag}_{{\bf k}\alpha}
c_{{\bf k}\alpha}+
J {\bf S}_{ei} {\bf S}_{fi}\, ,
\end{equation}
where ${\bf S}_{ei}$ and ${\bf S}_{fi}$ are the spin of the
conduction electrons and the local spin at site
$i$ respectively. As in many equations throughout
this paper, sum over repeated indexes is implicitly
understood. ${\bf S}_{ei}$ is given in terms of the
electronic operators in the site (Wannier) representation by
\begin{equation}
{\bf S}_{ei}=\frac{1}{2} c^{\dag}_{i\alpha}{\bf \sigma}_{\alpha
\beta} c_{i\beta}\, .
\end{equation}

If ${\bf R}_{i}$ is the position of the $i$-site, the
$c^{\dag}_{i\alpha}$-operators are given, for a lattice with $N$
sites, by
\begin{equation}
c^{\dag}_{i\alpha}=N^{-1/2}\sum_{{\bf k}}{\rm e}^{-i
{\bf R}_i{\bf k}} c^{\dag}_{{\bf k}\alpha}\, .
\end{equation}

The fundamental operators $c_{{\bf k}\alpha}$, ${\bf S}_{fi}$ are
characterized by the following algebraic properties:
\begin{eqnarray}
\left \{ c_{{\bf k}\alpha},c^{\dag}_{{\bf k'}\beta} \right \}&=&
\delta_{{\bf k} {\bf k'}}\delta_{\alpha\beta}\; , \;
\left \{ c_{{\bf k}\alpha},c_{{\bf k'}\beta} \right \} =0\, ,
\label{anticomrel} \\
\left[ S^{\alpha}_{fi},S^{\beta}_{fj} \right] &=&i \delta_{ij}
\epsilon_{\alpha\beta\gamma}
S^{\gamma}_{fi}\, , \label{comrel} \\
{\bf S}^{\dag}_{fi}&=&{\bf S}_{fi}\, ,\label{selfad} \\
{\bf S}^2_{fi}&=&3/4,\label{ssquare} \\
\left[ c_{{\bf k}\alpha},{\bf S}_{fi} \right]
&=&0\, . \label{comcs}
\end{eqnarray}

As we said before, the fundamental idea of this paper is a novel
treatment of the local moments: We shall first consider these
degrees of freedom only, adding the conduction electrons at the
end.

We can satisfy (\ref{comrel}) and (\ref{selfad}) by writing
\begin{equation}
{\bf S}_{fi}=\frac{1}{2} f^{\dag}_{i\alpha}{\bf \sigma}_{\alpha
\beta} f_{i\beta}\, , \label{sinf}
\end{equation}
where $f^{\dag}_{i\alpha}$ and $f_{i\alpha}$ are creation and
annihilation operators of $f$-electrons satisfying cannonical
anticommutation relations. From Eq.\ (\ref{sinf}) we get
\begin{equation}
{\bf S}^{2}_{fi}=\frac{3}{4} f^{\dag}_{i\alpha}f_{i\alpha}+
\frac{3}{2}f^{\dag}_{i\uparrow}f^{\dag}_{i\downarrow}
f_{i\uparrow}f_{i\downarrow}\, ,
\end{equation}
which implies that ${\bf S}^{2}_{fi}=0$ if the $i$-site is empty
or doubly occupied and ${\bf S}^{2}_{fi}=3/4$ if it is singly
occupied. Therefore, as it is physically obvious, the condition
(\ref{ssquare}) is equivalent to demanding that each site be
occupied by one and only one $f$-electron.

Thus, the Hilbert space in which the ${\bf S}_{fi}$-operators
act, should be the set of all vectors of the form
\begin{equation}
\sum_{\alpha_{i}=\uparrow , \downarrow} C_{\alpha_{1},\ldots,
\alpha_{N}} f^{\dag}_{1\alpha_{1}} f^{\dag}_{2\alpha_{2}}
\ldots f^{\dag}_{N \alpha_{N}} |0\rangle . \label{fspace}
\end{equation}

Since the spin operators ${\bf S}_{fi}$ preserve the number of
$f$-electrons at each site, they are well-defined in this space.
It will be convenient for future developments, however, to work
with the following equivalent set of operators:
\begin{eqnarray}
s_{1,i} & \equiv & \sqrt{2} \left( S^{x}_{fi}+i
S^{y}_{fi} \right)
= \sqrt{2} f^{\dag}_{i\uparrow}
f_{i\downarrow}\, , \nonumber \\
s_{0,i} & \equiv & 2 S^{z}_{fi}=
f^{\dag}_{i\uparrow} f_{i\uparrow}
-f^{\dag}_{i\downarrow} f_{i\downarrow}\, , \label{smalls} \\
s_{-1,i} & \equiv & \sqrt{2}
\left( S^{x}_{fi}-i S^{y}_{fi} \right)
= \sqrt{2} f^{\dag}_{i\downarrow} f_{i\uparrow}\, . \nonumber
\end{eqnarray}

The initial ${\bf S}_{fi}$-operators can always be recovered using
the inverse transformation:
\begin{equation}
S^{x}_{fi}=\frac{1}{2\sqrt{2}} (s_{1,i}+s_{-1,i})\; ,\;
S^{y}_{fi}=\frac{-i}{2\sqrt{2}} (s_{1,i}-s_{-1,i})\; ,\;
S^{z}_{fi}=\frac{1}{2} s_{0,i}\, . \label{bigs}
\end{equation}

It is straightforward to see that the operators (\ref{smalls})
acting on the space (\ref{fspace}) satisfy the following
multiplication table:
\begin{equation}
\begin{array}{rlrl}
s_{0,i} s_{1,i}&=s_{1,i}&\;\;\;\;\;\;\;\;
s_{0,i}s_{-1,i}&=-s_{-1,i} \\
s_{1,i} s_{0,i}&=-s_{1,i}&      s_{-1,i}s_{0,i}&=s_{-1,i} \\
s_{1,i} s_{-1,i}&=1+s_{0,i}&   s^{2}_{1,i}&=s^{2}_{-1,i}=0 \\
s_{-1,i} s_{1,i}&=1-s_{0,i}&   s^{2}_{0,i}&=1
\end{array}
\label{mtable}
\end{equation}
as well as the properties
\begin{equation}
s^{\dag}_{1,i}=s_{-1,i}\; , \; s^{\dag}_{0,i}=
s_{0,i}\, , \label{adjoints}
\end{equation}
\begin{equation}
\left[ s_{l,i},s_{l',j} \right]=0 \;\;\; \forall i\neq j\, .
\label{comij}
\end{equation}

Eqs.\ (\ref{mtable})--(\ref{comij}) can also be regarded as
the defining conditions of the algebra since, as it
can be readily checked, they imply Eqs.\
(\ref{comrel})--(\ref{ssquare}).

Let us now think in more physical terms: Since the local
moments do not interact among themselves, all the spin
configurations will have the same energy when the
conduction electrons are absent. Thus, the physical state
of the system can be regarded, in this situation, as a
random linear combination of the $2^{N}$ possible spin
configurations, namely,
\begin{equation}
|RVS\rangle =\frac{1}{\sqrt{2^{N}}} \sum_{\alpha_{i}=\uparrow,
\downarrow} {\rm e}^{i R_{\alpha_{1} \ldots
\alpha_{N}}} f^{\dag}_{1\alpha_{1}} \ldots
f^{\dag}_{N\alpha_{N}}|0\rangle ,
\end{equation}
where the phases $R_{\alpha_{1} \ldots \alpha_{N}}$ are
taken at random. We will call this state the {\em random
vacuum state\/} (RVS).

We can now apply to this state our fundamental
$s_{l,i}$-operators. As it is obvious from Eq.\ (\ref{smalls}),
their effect is to excite spin fluctuations in the vacuum
state. $s_{1,i}$ annihilates all the components with spin up
at site $i$ and flips the $i$-spin of the others; the result
is a normalized state with the $i$-spin up. $s_{-1,i}$ does
exactly the contrary and $s_{0,i}$ introduces a relative
phase shift of $\pi$ between the $i$-spin up and down
components.

By repeatedly applying the operators $s_{l,j}$ to the RVS,
we generate new states like, for example,
\begin{equation}
|\psi\rangle =s_{1,2} s_{0,5} s_{1,15} |RVS\rangle .
\end{equation}

We shall loosely say that $|\psi\rangle $ has spin fluctuations at
sites 2, 5, and 15. When applying $s_{l,i}$ to a state that
already has in its expression an operator in the same lattice
site, we can use the multiplication table (\ref{mtable})
to produce a state with one $s_{l,i}$-operator at most. For
example,
\begin{equation}
s_{-1,15}|\psi\rangle =s_{1,2} s_{0,5}|RVS\rangle
-s_{1,2}s_{0,5}s_{0,15}
|RVS\rangle .
\end{equation}

We can construct $4^{N}$ different combinations of spin
fluctuations by leaving each site either empty or with one of
the three possible fluctuations. The linear combinations of
these states form a vector space in which the $s_{l,j}$-operators
are perfectly defined through Eqs.\ (\ref{mtable})
and (\ref{comij}).

Using the randomness of the vacuum state, it is straightforward
to prove that the $4^{N}$ basic states constructed above are
orthonormal {\em in the thermodynamic limit\/}. For example, the
scalar product of $s_{0,1}|RVS\rangle $ and
$s_{1,1}|RVS\rangle $ is given by
\begin{equation}
\frac{1}{\sqrt{2}}\frac{1}{2^{N-1}}\sum_{\alpha_{2},\ldots,
\alpha_{N}=\uparrow,\downarrow}
{\rm e}^{-i R_{\uparrow \alpha_{2}
\ldots \alpha_{N}}} {\rm e}^{i R_{\downarrow \alpha_{2} \ldots
\alpha_{N}}}\, ,
\end{equation}
which is essentially the average of $2^{N-1}$ random phases and,
therefore, vanishes as $N \to \infty$. The result for the general
case can be obtained in an analogous way.

It should be emphasized that the exact orthonormality holds only
in the thermodynamic limit ($N \to \infty$). Actually, for $N$
finite, it would be mathematically impossible to have a set of
$4^{N}$ exactly orthonormal states in a vector space of dimension
$2^{N}$.

At this point, the physical ideas introduced in this section and
in particular the concept of RVS can be regarded as guiding
physical tools in the construction of our new algebraic structure.
Now that this structure has been established, we can completely
ignore the concept of RVS and present, as we do next, the obtained
representation of the algebra of the local moments in an axiomatic,
mathematically rigorous way:
\begin{enumerate}
\item Vector space: It is the set of all states of the form
\begin{equation}
\sum_{l_{i}=-1,0,1,2} C_{l_1 l_2 \ldots l_N}u_{l_1,1}u_{l_2,2}
\ldots u_{l_N,N} |\Phi\rangle ,
\end{equation}
where $C_{l_1 l_2 \ldots l_N} \in {\cal C}$, $u_{2,i}=1$, and
$u_{l_i,i}=s_{l_i,i}\,$ for $\, l_i=-1,0,1$.
\item Operators $s_{l,i}\,$: Their action on the vector space is
given by the multiplication table (\ref{mtable}) and Eq.\
(\ref{comij}).
\item Scalar product: It is defined by declaring that the basis
of the space formed by the $4^N$ vectors
$u_{l_1,1}u_{l_2,2}\ldots u_{l_N,N}|\Phi\rangle $ is orthonormal.
\end{enumerate}

These three points define a Hilbert space with operators $s_{l,i}$
satisfying Eqs.\ (\ref{mtable})--(\ref{comij})
or, equivalently, with operators ${\bf S}_{fi}$
[given by Eq.\ (\ref{bigs})] satisfying the
defining conditions of the algebra (\ref{comrel})--(\ref{ssquare}).

The inclusion of the conduction electrons is straightforward: The
basis for the complete Hilbert space is obtained by applying
electronic creation operators to the basis of spin fluctuations
presented above, and the action of the operators
$c^{\dag}_{{\bf k}\alpha}$, $c_{{\bf k}\alpha}$ on the states is
given by the cannonical anticommutation
relations (\ref{anticomrel}),
Eq.\ (\ref{comcs}) (where we can substitute ${\bf S}_{fi}$ by
$s_{l,i}$), and the condition $c_{{\bf k}\alpha}|\Phi\rangle =0\,$
$\forall {\bf k},\alpha$.

Finally, the Hamiltonian is given, in terms of the basic operators
of this new representation, by:
\begin{equation}
H=\varepsilon_{{\bf k}} c^{\dag}_{{\bf k}\alpha} c_{{\bf k}\alpha}
+\frac{J}{4} \left [ (c^{\dag}_{i\uparrow}c_{i\uparrow}-
c^{\dag}_{i\downarrow}c_{i\downarrow}) s_{0,i}+\sqrt{2}
c^{\dag}_{i\uparrow}c_{i\downarrow}s_{-1,i}+\sqrt{2}
c^{\dag}_{i\downarrow}c_{i\uparrow}s_{1,i} \right ].
\end{equation}

\section {Translational and spin rotational symmetries}

In this section, we determine the translational
and spin rotational
symmetries of the system within this new representation.
These symmetries will be used in the following sections to
characterize the spectrum of excitations of the system.
A unitary operator is completely determined by specifying
how it transforms the basic operators
$c^{\dag}_{{\bf k}\alpha}$, $s_{l,i}$ and its action on the
vacuum state: this is what we do next to introduce the
symmetries.

\subsection {Lattice translations}

It is physically evident that, if ${\bf a}_\gamma$ ($\gamma=
1,2,3$) are three primitive vectors of the real lattice,
their associated translation operators $T_{{\bf a}_\gamma}$
should satisfy:
\begin{eqnarray}
T_{{\bf a}_\gamma} c^{\dag}_{{\bf R}_i,\alpha}
T^{\dag}_{{\bf a}_\gamma}&=&c^{\dag}_{{\bf R}_i+{\bf a}_\gamma,
\alpha}\, , \\
T_{{\bf a}_\gamma} s_{l,{\bf R}_i}
T^{\dag}_{{\bf a}_\gamma}&=&s_{l,{\bf R}_i+{\bf a}_\gamma}\, ,\\
T_{{\bf a}_\gamma}|\Phi\rangle &=&|\Phi\rangle .
\end{eqnarray}

These equations define unitary operators that leave the
Hamiltonian invariant and thus represent the basic
translational symmetries of the system.

To exploit this symmetry it is convenient to Fourier transform
all operators to the momentum representation:
\begin{eqnarray}
c^{\dag}_{{\bf k}\alpha}&=&N^{-1/2} \sum_i {\rm e}^{i {\bf R}_i
{\bf k}} c^{\dag}_{i\alpha}\, , \\
s_{l,{\bf k}}&=&N^{-1/2} \sum_i {\rm e}^{i {\bf R}_i
{\bf k}} s_{l,i}\, ,
\end{eqnarray}
where ${\bf k}$ is always in the first Brillouin zone. Lattice
translations just multiply these operators by a phase factor:
\begin{eqnarray}
T_{{\bf a}_\gamma} c^{\dag}_{{\bf k}\alpha}
T^{\dag}_{{\bf a}_\gamma}&=&{\rm e}^{-i {\bf k} {\bf a}_\gamma}
c^{\dag}_{{\bf k}\alpha}\, , \\
T_{{\bf a}_\gamma} s_{l,{\bf k}}
T^{\dag}_{{\bf a}_\gamma}&=&{\rm e}^{-i {\bf k} {\bf a}_\gamma}
s_{l,{\bf k}}\, .
\end{eqnarray}

The $s_{l,{\bf k}}$ operators satisfy the following equations,
which will be of much use in the next sections:
\begin{eqnarray}
\left [ s_{0,{\bf k}},s_{\pm 1,
{\bf k}'} \right] &=&\pm 2 N^{-1/2}
s_{\pm 1,{\bf k}+{\bf k}'}\, , \\
\left [ s_{1,{\bf k}},s_{-1,{\bf k}'} \right] &=&2 N^{-1/2}
s_{0,{\bf k}+{\bf k}'}\, , \\
\left [ s_{l,{\bf k}},s_{l,{\bf k}'} \right] &=&0\, , \\
s^{\dag}_{l,{\bf k}}&=&s_{-l,-{\bf k}}\, .
\end{eqnarray}

Finally, the Hamiltonian is given, in the momentum
representation, by
\begin{equation}
H=\varepsilon_{{\bf k}} c^{\dag}_{{\bf k}\alpha} c_{{\bf k}\alpha}
+\frac{J}{4 N^{1/2}} \left [ (c^{\dag}_{{\bf k}\uparrow}
c_{{\bf k}'\uparrow}-c^{\dag}_{{\bf k}\downarrow}
c_{{\bf k}'\downarrow}) s_{0,{\bf k}'-{\bf k}}+ \sqrt{2}
c^{\dag}_{{\bf k}\uparrow} c_{{\bf k}'\downarrow}
s_{-1,{\bf k}'-{\bf k}}+\sqrt{2} c^{\dag}_{{\bf k}\downarrow}
c_{{\bf k}'\uparrow}s_{1,{\bf k}'-{\bf k}} \right] .
\end{equation}

\subsection {Spin rotations}

The spin rotational symmetry is quite subtle. The total-spin
operators of the system are given by
\begin{equation}
{\bf \Sigma}=\sum_i {\bf S}_{ei}+{\bf S}_{fi}\, ,
\end{equation}
and satisfy
\begin{eqnarray}
\left [ \Sigma^\alpha,\Sigma^\beta \right] &=& i
\epsilon_{\alpha\beta\gamma}\Sigma^\gamma, \label{ss}\\
{\bf \Sigma}^{\dag}&=&{\bf \Sigma}\, ,\label{splus}\\
\left[ H,{\bf \Sigma} \right] &=&0. \label{hs}
\end{eqnarray}

These are the properties that characterize the generators of
a unitary representation of $SU(2)$ that leaves the Hamiltonian
invariant. There is, however, an important problem: the
operators ${\bf \Sigma}$ {\em do not\/} annihilate the vacuum state
and, therefore, the unitary transformations generated by them
would not leave $|\Phi\rangle $ invariant. Since (in the absence of
magnetic order) we do not physically expect the spin rotational
symmetry to be broken, ${\bf \Sigma}$ cannot be taken as the
generators of this symmetry. We should thus look for another set
of operators ${\bf S}$ satisfying Eqs.\ (\ref{ss})--(\ref{hs}) plus
the condition
\begin{equation}
{\bf S} |\Phi\rangle =0\, .
\label{sannihilates}
\end{equation}

As we already saw, every vector in the Hilbert space constructed
in Section II can be written as $V|\Phi\rangle $,
where $V$ is a linear
combination of products of operators $c^{\dag}_{i\alpha}$ and
$s_{l,j}$. The effect of a transformation $\exp(i{\bf \alpha}
{\bf \Sigma})$ on a state would be
\begin{equation}
{\rm e}^{i{\bf \alpha}{\bf \Sigma}}V|\Phi\rangle =
{\rm e}^{i{\bf \alpha}{\bf \Sigma}}V
{\rm e}^{-i{\bf \alpha}{\bf \Sigma}}
\left[ {\rm e}^{i{\bf \alpha}{\bf \Sigma}}|\Phi\rangle  \right],
\end{equation}
where
\begin{equation}
{\rm e}^{i{\bf \alpha}{\bf \Sigma}}|\Phi\rangle \neq|\Phi\rangle .
\end{equation}

The obvious thing to do to satisfy Eqs.\ (\ref{ss})--(\ref{hs})
plus the condition (\ref{sannihilates}), is to demand that $\exp
(i{\bf \alpha}{\bf S})$ transform $V$ in the same way as
$\exp(i{\bf \alpha}{\bf \Sigma})$ but leave
$|\Phi\rangle $ invariant, namely,
\begin{equation}
{\rm e}^{i{\bf \alpha}{\bf S}}V|\Phi\rangle =
{\rm e}^{i{\bf \alpha}{\bf \Sigma}}V
{\rm e}^{-i{\bf \alpha}{\bf \Sigma}}|\Phi\rangle ,
\end{equation}
which implies that the action of ${\bf S}$ on the Hilbert
space is given by
\begin{equation}
{\bf S} V|\Phi\rangle =\left[ {\bf \Sigma},V \right] |\Phi\rangle.
\label{defs}
\end{equation}

It is quite straightforward to prove that the linear operators
${\bf S}$ {\em defined\/} by Eq.\ (\ref{defs}) satisfy Eqs.\
(\ref{ss})--(\ref{sannihilates}) and will, therefore, be taken as
the generators of the spin rotational symmetry of the system.
This implies, in particular, that the conduction electrons
$c^{\dag}_{{\bf k}\alpha}|\Phi\rangle $
($\alpha=\uparrow,\downarrow$) and the spin fluctuations
$s_{l,{\bf k}}|\Phi\rangle $ ($l=-1,0,1$) form,
respectively, an $s=1/2$
doublet and an $s=1$ triplet under spin rotations, as is
physically expected.

\section {Kondo lattice with one conduction electron}

We shall consider, in this section, the case of a single
conduction electron interacting with the lattice of
local moments. This system is interesting because it is
simpler than the general case and reveals very clearly
the appearance of heavy fermions as collective states
of the bare conduction electron and spin fluctuations.

We shall first examine the atomic limit (band
width$\to 0$) to determine that the spectrum of the
system consists in two $s=1/2$ doublet bands and an
$s=3/2$ quadruplet. Then, we will study the structure
of these bands away from the atomic limit using the
lowest approximation in which the conduction electron
can only dress with one spin fluctuation at most.
Finally, we will see that this study
reveals the formation of collective, heavy-fermion
states for low relative values of the exchange coupling.

The KLM is particularly simple in the atomic limit; it
is just the sum of local Hamiltonians associated to each
lattice site:
\begin{equation}
H=\sum_i H_i \;\; \left( H_i=J {\bf S}_{ei}
{\bf S}_{fi} \right).
\end{equation}

Thus, in this limit, the one-electron problem reduces to
diagonalizing $H_i$ in the space of the local degrees of
freedom with a single electron present. In the
representation introduced in Section II, this space is
generated by the following eight vectors:
\begin{eqnarray}
c^{\dag}_{i\alpha}|\Phi\rangle &\;\;\;\alpha=\uparrow,
\downarrow\, ,\label{ce}\\
c^{\dag}_{i\alpha}s_{l,i}|\Phi\rangle &\;\;\;\alpha=\uparrow,
\downarrow\;,\; l=-1,0,1\, ; \label{cesf}
\end{eqnarray}
(a state with more than one spin fluctuation accompanying
the conduction electron at site $i$ can be developed by
the multiplication table (\ref{mtable}) into a linear
combination of these states).

Since the spin operators ${\bf S}$ are well-defined in this
local space and commute with $H_i$, they can be used to
diagonalize the local Hamiltonian. The diagonalization of
${\bf S}^2$ is straightforward: the vectors (\ref{ce})
form an $s=1/2$ doublet while the six vectors (\ref{cesf})
are the product of an $s=1/2$ by an $s=1$ representations
and can be decomposed into an $s=1/2$ doublet and an
$s=3/2$ quadruplet. The energy of the quadruplet is seen to
be $J/4$ while the diagonalization of $H_i$ in the subspace
of $s=1/2$ leads to two doublets with energies $J/4$ and
$-3J/4$. The states in each multiplet can be determined,
for example, by repeatedly applying the lowering operator
($S^{-}=S^{x}-iS^{y}$) to the states with the highest
eigenvalue of $S^{z}$ ($s^{z}=s$). The explicit expressions
of these vectors are given below.
\begin{enumerate}
\item Quadruplet ($E=J/4$):
\begin{eqnarray}
|\frac{J}{4},\frac{3}{2},\frac{3}{2},i\rangle
&=&c^{\dag}_{i\uparrow}
s_{1,i}|\Phi\rangle ,\label{quadru} \\
|\frac{J}{4},\frac{3}{2},\frac{1}{2},i\rangle
&=&\frac{1}{\sqrt{3}}
S^{-}|\frac{J}{4},\frac{3}{2},\frac{3}{2},i\rangle =
\frac{1}{\sqrt{3}}(c^{\dag}_{i\downarrow}s_{1,i}-\sqrt{2}
c^{\dag}_{i\uparrow}s_{0,i})|\Phi\rangle , \\
|\frac{J}{4},\frac{3}{2},\frac{-1}{2},i\rangle &=&
-\frac{1}{2}S^{-}|\frac{J}{4},\frac{3}{2},\frac{1}{2},i\rangle =
\frac{1}{\sqrt{3}}(c^{\dag}_{i\uparrow}s_{-1,i}+\sqrt{2}
c^{\dag}_{i\downarrow}s_{0,i})|\Phi\rangle , \\
|\frac{J}{4},\frac{3}{2},\frac{-3}{2},i\rangle
&=&\frac{1}{\sqrt{3}}
S^{-}|\frac{J}{4},\frac{3}{2},\frac{-1}{2},i\rangle =
c^{\dag}_{i\downarrow}s_{-1,i}|\Phi\rangle .
\end{eqnarray}
\item $E=J/4$ doublet:
\begin{eqnarray}
|\frac{J}{4},\frac{1}{2},\frac{1}{2},i\rangle
&=&\frac{\sqrt{3}}{6}
(3c^{\dag}_{i\uparrow}+c^{\dag}_{i\uparrow}s_{0,i}+\sqrt{2}
c^{\dag}_{i\downarrow}s_{1,i})|\Phi\rangle ,\\
|\frac{J}{4},\frac{1}{2},\frac{-1}{2},i\rangle &=&
S^{-}|\frac{J}{4},\frac{1}{2},\frac{1}{2},i\rangle
=\frac{\sqrt{3}}{6}
(3c^{\dag}_{i\downarrow}-c^{\dag}_{i\downarrow}s_{0,i}+\sqrt{2}
c^{\dag}_{i\uparrow}s_{-1,i})|\Phi\rangle .
\end{eqnarray}
\item $E=-3J/4$ doublet:
\begin{eqnarray}
|\frac{-3J}{4},\frac{1}{2},\frac{1}{2},i\rangle &=&\frac{1}{2}
(c^{\dag}_{i\uparrow}-c^{\dag}_{i\uparrow}s_{0,i}-\sqrt{2}
c^{\dag}_{i\downarrow}s_{1,i})|\Phi\rangle ,\\
|\frac{-3J}{4},\frac{1}{2},\frac{-1}{2},i\rangle &=&
S^{-}|\frac{-3J}{4},\frac{1}{2},\frac{1}{2},i\rangle =\frac{1}{2}
(c^{\dag}_{i\downarrow}+c^{\dag}_{i\downarrow}s_{0,i}-\sqrt{2}
c^{\dag}_{i\uparrow}s_{-1,i})|\Phi\rangle .
\end{eqnarray}
\end{enumerate}

Thus, the Fourier transform of these vectors, namely,
\begin{equation}
|E,s,s^z,{\bf k}\rangle =N^{-1/2}\sum_i
{\rm e}^{i {\bf R}_i{\bf k}}
|E,s,s^z,i\rangle  \label{fourier}
\end{equation}
(with ${\bf k}$ in the first Brillouin zone), are exact
eigenstates of the operators $H$, ${\bf S}^2$, $S^z$, and
$T_{{\bf a}_\gamma}$, in the atomic limit, with eigenvalues
$E$, $s(s+1)$, $s^z$, and $\exp(-i{\bf k}{\bf a}_\gamma)$
respectively. These operators form a complete set in the sense
that each vector is characterized by a definite set of
eigenvalues. Since ${\bf S}^2$, $S^z$, and $T_{{\bf a}_\gamma}$
{\em always\/} commute with $H$, they can be used to characterize
the spectrum of the system away from the atomic limit as well.
This simplifies enormously the calculation of the spectrum:
we just have to diagonalize $H$ in the subspace of vectors with
definite eigenvalues of these operators.

Thus, from the analysis of the atomic limit and the symmetries
of the system, we conclude that the spectrum contains two
$s=1/2$ doublet bands and an $s=3/2$ quadruplet. We shall
now propose how to calculate these bands away from the
atomic limit. Since all the members in a spin multiplet have
the same energy, we only need to calculate the energy of
one of them; we shall select the one with $s^z=s$.

The most general state of the one-electron
system is a linear combination
of terms containing a single $c^{\dag}_{{\bf k}\alpha}$-operator
multiplied by several spin fluctuations corresponding to various
lattice sites. The approximation that we propose to calculate
the spectrum of excitations consists in determining variationally
the best approximation to the actual eigenstates of $H$ within
the subspace in which only a maximum of $n$ spin
fluctuations can accompany the conduction electron.

One should start by writing the most general vector in this
subspace with definite eigenvalues of $T_{{\bf a}_\gamma}$,
${\bf S}^2$, and $S^{z}$. We shall denote this state
(with $s^z=s$) by $|\xi,s,s,{\bf k}\rangle
=V_{\xi,s,{\bf k}}|\Phi\rangle $,
where $\xi$ represents the parameters of this general vector.
The diagonalization of $T_{{\bf a}_\gamma}$ is straightforward
if we consider all the degrees of freedom in momentum space,
and the spin eigenstate conditions are equivalent to
\begin{eqnarray}
S^{+}|\xi,s,s,{\bf k}\rangle
&=&\left[ \Sigma^{+},V_{\xi,s,{\bf k}}
\right] |\Phi\rangle =0\, ,\\
S^z|\xi,s,s,{\bf k}\rangle &=&\left[ \Sigma^{z},V_{\xi,s,{\bf k}}
\right] |\Phi\rangle =s |\xi,s,s,{\bf k}\rangle .
\end{eqnarray}

The action of $H$ on $|\xi,s,s,{\bf k}\rangle $ can be written as
\begin{equation}
H|\xi,s,s,{\bf k}\rangle =E_{\bf k}|\xi,s,s,{\bf k}\rangle
+|\perp\rangle , \label{apeigen}
\end{equation}
where $|\perp\rangle $ is a vector orthogonal
to $|\xi,s,s,{\bf k}\rangle $.
Obviously, the best approximation to the actual eigenstate will
be the normalized vector for which $\langle \perp|\perp\rangle $
is minimized.
Thus, the optimal $\xi$-parameters should be determined by
demanding that
\begin{equation}
\langle \perp|\perp\rangle =\langle \xi,s,s,{\bf k}|H^2|\xi,s,s,
{\bf k}\rangle -E^2_{\bf k} \label{ximin}
\end{equation}
is minimized, where $E_{\bf k}$
(the approximate energy) is given by
\begin{equation}
E_{\bf k}=\langle \xi,s,s,{\bf k}|H|\xi,s,s,{\bf k}\rangle ,
\label{xie}
\end{equation}
and $|\xi,s,s,{\bf k}\rangle $ satisfies the normalization
condition
\begin{equation}
\langle \xi,s,s,{\bf k}|\xi,s,s,{\bf k}\rangle =1\, .
\label{xinorm}
\end{equation}

In this article, we will consider only the lowest ($n=1$)
approximation. From Eqs.\ (\ref{quadru})--(\ref{fourier}) it is
clear that this approximation gives the exact result in the
atomic limit and it should, therefore, yield precise results
around this limit. It also describes exactly the bare electrons
when $J=0$ and it is thus expected to express accurately how
these excitations are modified when a small interaction is
introduced. Away from these situations, this approximation is
expected to describe the spectrum only qualitatively.

The determination of the states $|\xi,1/2,1/2,{\bf k}\rangle $
and $|\xi,3/2,3/2,{\bf k}\rangle $ in the lowest
approximation is straightforward, the result being
\begin{eqnarray}
|[\alpha,\beta],\frac{1}{2},\frac{1}{2},{\bf k}\rangle &=&
\left[ \alpha_{{\bf k}}c^{\dag}_{{\bf k}\uparrow}+N^{-1/2}
\beta_{{\bf k}}({\bf q})(c^{\dag}_{{\bf q}\uparrow}
s_{0,{\bf k}-{\bf q}}+\sqrt{2}c^{\dag}_{{\bf q}\downarrow}
s_{1,{\bf k}-{\bf q}}) \right] |\Phi\rangle ,\label{xivecd} \\
|[\gamma],\frac{3}{2},\frac{3}{2},{\bf k}\rangle &=&N^{-1/2}
\gamma_{{\bf k}}({\bf q})c^{\dag}_{{\bf q}\uparrow}
s_{1,{\bf k}-{\bf q}}|\Phi\rangle ,\label{xivecq}
\end{eqnarray}
where sum over ${\bf q}$ is implicitly understood. It is also
assumed that ${\bf k}-{\bf q}$ is in the first Brillouin zone,
which occasionally requires the addition of a reciprocal lattice
vector. $\alpha$, $\beta$, and $\gamma$ are the variational
parameters whose optimal values should be determined by the
conditions (\ref{ximin})--(\ref{xinorm}). The mathematical
development of these conditions is a lengthy technical task
which is outlined in the Appendix.

Since we will be considering the system in all the situations
between the atomic and $J=0$ limits, it is important to write
the Hamiltonian in terms of parameters that
scale in a convenient way. Since the number of conduction
electrons is a constant, we can shift the bare dispersion
relation $\varepsilon_{{\bf k}}$ so that
$\varepsilon_{{\bf k}}=\Delta \eta_{{\bf k}}$, where $\Delta$
is the band width and $\eta_{{\bf k}}$ is a normalized
band of unit width centered at $\varepsilon=0$ ($-1/2\leq
\eta_{{\bf k}}\leq 1/2$).

Instead of working with $J$, $\Delta$, and $\eta_{{\bf k}}$,
we will use the more convenient parameters $\lambda$,
$\Lambda$, and $\eta_{{\bf k}}$ given by
\begin{eqnarray}
\lambda&\equiv& \frac{J}{\left( \Delta+|J| \right)}\, ,\\
\Lambda&\equiv&\Delta+|J|\, .
\end{eqnarray}

$|\lambda|$ is a dimensionless parameter describing the
strength of the interaction and it varies from zero (no
interaction) to one (atomic limit). Positive and negative
values of $\lambda$ correspond, respectively, to antiferromagnetic
and ferromagnetic couplings. $\Lambda$ is just a measure of the
global energy scale of the system which factorizes in $H$, and
can thus be set equal to one without loss of generality. The
Hamiltonian, therefore, depends essentially on the parameters
$\lambda$ and $\eta_{{\bf k}}$ and is given by
\begin{equation}
H=\varepsilon_{{\bf k}} c^{\dag}_{{\bf k}\alpha}c_{{\bf k}\alpha}+
\lambda {\bf S}_{ei}{\bf S}_{fi},
\end{equation}
where $\varepsilon_{{\bf k}}=(1-|\lambda|)\eta_{{\bf k}}$,
$-1/2\leq\eta_{{\bf k}}\leq 1/2$, and $-1\leq\lambda\leq 1$.

As we said before, the development of the conditions
(\ref{ximin})--(\ref{xinorm}) that determine the optimal parameters
of the approximate eigenstates (\ref{xivecd}) and (\ref{xivecq}) is
carried out in the Appendix. We next summarize the results obtained
for the doublet and quadruplet bands:
\begin{enumerate}
\item Doublet bands: The optimal parameters $\alpha$ and
$\beta$ in Eq.\ (\ref{xivecd}) and the approximate energy $E$
are given by
\begin{eqnarray}
\alpha_{{\bf k}}&=&{\rm Re}\left[
4 a_{{\bf k}}A_{{\bf k}}\right],\\
\beta_{{\bf k}}({\bf q})&=&\frac{\lambda a_{{\bf k}}}
{\varepsilon_{{\bf q}}-\varepsilon_{{\bf k}}+w_{{\bf k}}}+
\frac{\lambda a^{\star}_{{\bf k}}}
{\varepsilon_{{\bf q}}-\varepsilon_{{\bf k}}+w^{\star}_{{\bf k}}}
\, ,\\
E_{{\bf k}}&=&\varepsilon_{{\bf k}}-{\rm Re}[w_{{\bf k}}]\, ,
\end{eqnarray}
where
\begin{eqnarray}
a^2&=&-\frac{w^{\star}A^{\star}}{w A}\frac{{\rm Im}[w^{\star 2}w]}
{{\rm Im}\left[ w^{\star}A^{\star}[8Aw^{\star}
w-(4A+3B/A)w^{\star}
(w^{\star}-w)-6\lambda w] \right]}\, ,\label{a2}\\
A_{{\bf k}}&=&\lambda \int^{1/2}_{-1/2} \frac{D(\eta)\,d\eta}
{(1-|\lambda|)\eta-\varepsilon_{{\bf k}}+w_{{\bf k}}}-2
\, ,\label{Aofk}\\
B_{{\bf k}}&=&\lambda^2 \int^{1/2}_{-1/2} \frac{D(\eta)\,d\eta}
{\left[ (1-|\lambda|)\eta-\varepsilon_{{\bf k}}+
w_{{\bf k}}\right]^2}\, ,\label{Bofk}
\end{eqnarray}
and $w_{{\bf k}}$ is the solution
of the following pair of equations
\begin{equation}
{\rm Im}\left[ w^{\star}(8w^{\star} w+3\lambda(1+2/A)w^{\star}-
3\lambda^2/2)\right]=0,\label{eqw1}
\end{equation}
\begin{equation}
{\rm Im}\left[w^{\star 2}A^{\star}\left[
(4A+3B/A)(w^{\star}-w)^2+
3\lambda (A+2)(w^{\star}-3w)+6\lambda(1-A/A^{\star})w\right]
\right]=0.\label{eqw2}
\end{equation}

$D(\eta)$ in Eqs.\ (\ref{Aofk}) and (\ref{Bofk}) represents the
density of states of the normalized band $\eta_{{\bf k}}$. The
subscripts ${\bf k}$ in Eqs.\ (\ref{a2}), (\ref{eqw1}), and
(\ref{eqw2}) have been omitted to simplify the notation.

\item Quadruplet band: The optimal $\gamma$ in Eq.\
(\ref{xivecq}) and the approximate energy $E$ are given by
\begin{equation}
\gamma ({\bf q})=\frac{a}{\varepsilon_{{\bf q}}-z}+
\frac{a^{\star}}{\varepsilon_{{\bf q}}-z^{\star}}\;\; ,\;\;
E={\rm Re}[z],
\end{equation}
where
\begin{eqnarray}
a^2&=&\frac{\lambda C^{\star}{\rm Im}[z]}
{2 C {\rm Im}\left[ C^{\star}(1+Dz/C)\right]}\, ,\\
C&=&\lambda \int^{1/2}_{-1/2} \frac{D(\eta)\,d\eta}
{(1-|\lambda|)\eta-z}+4\, ,\\
D&=&\lambda \int^{1/2}_{-1/2} \frac{D(\eta)\,d\eta}
{\left[ (1-|\lambda|)\eta-z\right]^2}\, ,
\end{eqnarray}
\noindent and $z$ is the simultaneous solution of
\begin{equation}
{\rm Im}\left[ C^2 D^{\star}\right]=0\; , \;
{\rm Im}\left[ C(zC^{\star}-3\lambda)\right]=0. \label{eqz}
\end{equation}
\end{enumerate}

It is clear from these equations that the energy $E_{{\bf k}}$
of the doublet bands depends on $\lambda$,
$\varepsilon_{{\bf k}}$,
and $D(\eta)$. In the case of the quadruplet, however,
$E_{{\bf k}}$ depends exclusively on $\lambda$ and $D(\eta)$,
which implies that this band is flat in this approximation.

In order to carry out explicit calculations, we must specify
$D(\eta)$. A realistic density of states should behave like
$\sqrt{1/2\pm\eta}$ at the extreme
points $\eta=\mp 1/2$ and must be
normalized. We shall take the simplest function satisfying
these conditions, namely,
\begin{equation}
D(\eta)=\frac{8}{\pi}\sqrt{\frac{1}{4}-\eta^2}\, ,
\end{equation}
which leads to simple expressions for A, B, C, and D
[see Eqs.\ (\ref{aint}) and (\ref{bint})].

At this point, we can already determine the lowest approximation
to the spectrum of the system by numerically solving Eqs.\
(\ref{eqw1}), (\ref{eqw2}) and (\ref{eqz}).

The solutions for the energies satisfy $E(-\lambda,
-\varepsilon)=-E(\lambda,\varepsilon)$
for the doublet bands, and $E(-\lambda)=
-E(\lambda)$ for the quadruplet. Since we can relate, with
these equations, the spectra for positive and negative values
of $\lambda$, we will restrict our study to the
antiferromagnetic case ($\lambda>0$).

The numerical results obtained are displayed in Fig.\ 1. In order
to make the physical interpretation of the results easier, we
have parameterized the values of the bare band $\varepsilon$ in
a band-like manner, namely, with the function $\varepsilon(\theta)=
-1/2(1-\lambda)\cos\theta$ ($\theta\in[0,\pi]$), which is
represented by the dashed curve. The solid curves correspond
to the solutions obtained for the two doublets $E(\lambda,
\varepsilon(\theta))$ and the quadruplet $E(\lambda)$.

Fig.\ 1(a) represents the atomic limit, with completely localized
bands at $E=1/4$ (quadruplet and upper doublet) and $E=-3/4$
(lower doublet). In Fig. 1(b), the degeneracy of the upper bands
in the atomic limit is broken; as we already mentioned, the
quadruplet band remains flat for all values of $\lambda$ in this
approximation. The transition from Fig. 1(c) to Fig. 1(d) is very
revealing: There is a clear tendency to form, with the states of
the superior and inferior parts of the upper and lower doublets
respectively, a band that closely follows the bare conduction
band, while the rest of the states tend to form a practically
flat band. We shall call these bands the {\em continuous\/}
(because it is essentially a continuation of the bare band) and
the {\em heavy-fermion\/} band respectively. Fig. 1(d) can actually
be interpreted as a hybridization of these two bands with the
characteristic pseudogap structure.

To get a deeper insight into the nature of these bands, we will
study the parameter $\alpha$. It is clear from Eq.\ (\ref{xivecd}),
that $\alpha^2_{{\bf k}}$ represents the probability for the
mode $|[\alpha,\beta],1/2,1/2,{\bf k}\rangle $
to be in the bare state
$c^{\dag}_{{\bf k}\uparrow}|\Phi\rangle $.
If the subscripts $U$ and
$L$ denote the upper and lower doublets respectively, we get, for
$\lambda=0.26$,
\begin{eqnarray}
\alpha^2_L(\theta=0)&=&0.77\; ,\;\alpha^2_U(\theta=\pi)=0.93,
\label{cont}\\
\alpha^2_U(\theta=0)&=&0.22\; ,\;\alpha^2_L(\theta=\pi)=0.018.
\label{hf}
\end{eqnarray}

The points in rows (\ref{cont}) and (\ref{hf}) correspond to the
continuous and heavy-fermion bands respectively. The interpretation
is clear: While the continuous band consists essentially in a
small modification of the bare electronic states, in the
heavy-fermion modes, the conduction electron strongly combines
with spin fluctuations forming a complex, collective state as
in the Kondo problem. It should be noted that the $s=3/2$
quadruplet states are also collective modes in the most strict
sense since, due to spin conservation, they have no probability
to be in the bare states.

Calculations for very small values of $\lambda$ show that the
continuous band actually coincides with the bare
band as $\lambda\to 0$. Since, as we already argued, these
calculations are very reliable, we can assert that there is no
hybridization of the continuous and heavy-fermion bands for
very small couplings. Figs.\ 1(f), 1(e), and 1(d), describe
the process of hybridization of these bands around $\lambda=
0.24$. We find that beyond the poins A, B, C, and D in
Figs.\ 1(e) and 1(f), the system of Eqs.\
(\ref{eqw1}) and (\ref{eqw2})
no longer has a solution. We have interpreted this fact as a
mathematical expression of the inadequacy of the lowest order
approximation around these points: one should physically
expect that the precise description of the collective states
will require considering more than just a single spin
fluctuation combining with the bare conduction electron.

\section{Summary and conclusions}

With the introduction of a new representation of the algebra
of the local moments, we have established a new framework for
the study of the Kondo lattice which enables a very clear
mathematical description of heavy fermions as collective states
of conduction electrons and spin fluctuations of the lattice
of local moments.

\section*{Acknowledgements}
This work has been financed by the DGICYT (Research
Project No.\ PB93-1249).
J.M.~Prats acknowledges financial support from a postdoctoral
fellowship from the {\em Ministerio de Educacion y Ciencia\/} of
Spain.

\appendix
\section*{Lowest order equations for the doublet bands}

We shall briefly develop here the conditions
(\ref{ximin})--(\ref{xinorm})
for determining the optimal parameters $\alpha$ and $\beta$ that
characterize the doublet bands of the one-electron Kondo lattice
in the lowest approximation. The development of the corresponding
equations for the quadruplet band is completely analogous and
will be omitted.

Let us summarize the problem: We have to determine the parameters
$\alpha$ and $\beta$ of the vector
$|[\alpha,\beta],1/2,1/2,{\bf k}\rangle $ in Eq.\
(\ref{xivecd}) that minimize the function
\begin{equation}
\Psi_{{\bf k}}\left( [\alpha,\beta]\right)=
\langle [\alpha,\beta],\frac{1}{2},\frac{1}{2},{\bf k}|H^2
|[\alpha,\beta],\frac{1}{2},\frac{1}{2},{\bf k}\rangle
-E^2_{{\bf k}}\, ,
\end{equation}
where
\begin{equation}
E_{{\bf k}}=
\langle [\alpha,\beta],\frac{1}{2},\frac{1}{2},{\bf k}|H
|[\alpha,\beta],\frac{1}{2},\frac{1}{2},{\bf k}\rangle ,
\end{equation}
while satisfying the normalization condition
\begin{equation}
\langle [\alpha,\beta],\frac{1}{2},\frac{1}{2},{\bf k}|[\alpha,
\beta], \frac{1}{2},\frac{1}{2},{\bf k}\rangle =1.
\label{normcond}
\end{equation}

In order to handle this normalization condition, it is very
convenient to introduce a Lagrange multiplier $\tau$. Thus, the
objective will be to find the parameters $\alpha$ and $\beta$
that minimize
\begin{equation}
\Omega_{{\bf k}}\left( [\alpha,\beta],\tau\right)=
\Psi_{{\bf k}}\left( [\alpha,\beta]\right)+
\tau \langle [\alpha,\beta],\frac{1}{2},\frac{1}{2},{\bf k}
|[\alpha,\beta],
\frac{1}{2},\frac{1}{2},{\bf k}\rangle ,
\label{omega}
\end{equation}
for an arbitrary $\tau$, and then select the value of $\tau$ for
which the normalization condition (\ref{normcond}) is satisfied.

To simplify the notation, we will often omit the subindexes
${\bf k}$ in the forthcoming equations. If $m_1$, $m_2$,
$n_1$, and $n_2$ are defined by
\begin{equation}
\begin{array}{rlrl}
m_1&=\frac{1}{\Gamma}\int \beta({\bf q})\,d{\bf q}
\, ,&m_2=&\frac{1}{\Gamma}\int\beta^{\star}({\bf q})\beta({\bf q})\,
d{\bf q}\, ,\\
n_1&=\frac{1}{\Gamma}\int\varepsilon_{{\bf q}}
\beta({\bf q})\,d{\bf q}\, ,\;\;&n_2=&
\frac{1}{\Gamma}\int\varepsilon_{{\bf q}}\beta^{\star}
({\bf q})\beta({\bf q})\,d{\bf q}\, ,
\end{array}
\end{equation}
where the integrals are extended over the first Brillouin zone and
$\Gamma$ is the volume of this zone, a lengthy calculation yields
\begin{equation}
\langle [\alpha,\beta],\frac{1}{2},\frac{1}{2},{\bf k}
|[\alpha,\beta], \frac{1}{2},\frac{1}{2},{\bf k}\rangle
=\alpha^{\star}\alpha+3m_2\, ,
\end{equation}
\begin{equation}
E=\alpha^{\star}\alpha\varepsilon+\frac{3\lambda}{4} (\alpha
m_1^{\star}+\alpha^{\star}m_1)-\frac{3\lambda}{2}m_1^{\star}
m_1+3n_2\, ,
\end{equation}
\begin{eqnarray}
\langle [\alpha,\beta],\frac{1}{2},\frac{1}{2},
{\bf k}|H^2|[\alpha, \beta],\frac{1}{2},
\frac{1}{2},{\bf k}\rangle & &=\frac{3}{\Gamma}
\int\varepsilon^2_{{\bf q}}\beta^{\star}({\bf q})\beta({\bf q})\,
d{\bf q}+\frac{3\lambda n_1^{\star}}{2}(\alpha/2-m_1)+\nonumber\\
& &\frac{3\lambda n_1}{2}(\alpha^{\star}/2-m^{\star}_1)+
\frac{9}{16}\lambda^2m_2+\frac{3}{4}\lambda^2m_1^{\star}m_1+
\nonumber\\
& &\frac{3}{4}\lambda(\varepsilon-\lambda/2)(\alpha
m_1^{\star}+\alpha^{\star}m_1)+(\varepsilon^2+3\lambda^2/16)
\alpha^{\star}\alpha\, .
\end{eqnarray}

Substituting these equations in (\ref{omega}), we can determine
$\Omega_{{\bf k}}\left( [\alpha,\beta],\tau\right)$ and
calculate its variations with respect to $\beta({\bf q})$
and $\alpha$. The equations that express the vanishing of
these variations are, respectively,
\begin{eqnarray}
0&=&\beta({\bf q})\left[ \varepsilon^2_{{\bf q}}
-2E\varepsilon_{{\bf q}}+\frac{3\lambda^2}{16}+
\tau\right] +\frac{\lambda}{2}(\alpha/2-m_1)
\varepsilon_{{\bf q}}+\frac{\lambda^2m_1}{4}+ \nonumber\\
& &\frac{\lambda\alpha}{4}(\varepsilon-\lambda/2)-\frac{\lambda
n_1}{2}-\lambda E(\alpha/2-m_1)\;\;\forall{\bf q}
\, ,\label{betastr}\\
0&=&\alpha(\varepsilon^2+3\lambda^2/16+\tau-2E\varepsilon)+
\frac{3\lambda}{4}\left[ n_1+m_1(\varepsilon-
\lambda/2-2E)\right] \, .\label{varalpha}
\end{eqnarray}

We see from Eq.\ (\ref{betastr}), that the function
$\beta({\bf q})$ has the structure
\begin{equation}
\beta({\bf q})=\frac{\lambda a}{\varepsilon_{{\bf q}}-z}+
\frac{\lambda b}{\varepsilon_{{\bf q}}-z^{\star}}\, ,
\end{equation}
where the complex parameters $a$, $b$, and $z$ satisfy
\begin{eqnarray}
a+b&=&-\frac{\alpha}{4}+\frac{m_1}{2}\, ,\label{first}\\
-zb-z^{\star}a&=&\frac{n_1}{2}-\frac{m_1}{2}
(z+z^{\star}+\lambda/2)
+\frac{\alpha}{4}(z+z^{\star}+\lambda/2-\varepsilon)\, ,\\
z+z^{\star}&=&2E=2\alpha\alpha^{\star}\varepsilon+
\frac{3\lambda}{2}
(\alpha m_1^{\star}+\alpha^{\star}m_1)-3\lambda m_1^{\star}m_1
+6n_2\, ,\\
z^{\star}z&=&\frac{3\lambda^2}{16}+\tau\, .\label{last}
\end{eqnarray}

A complete set of equations is formed by Eqs.\
(\ref{first})--(\ref{last}) plus Eq.\ (\ref{varalpha}),
which can be written as
\begin{equation}
\alpha(\varepsilon-z)(\varepsilon-z^{\star})+\frac{3\lambda}{4}
\left[ n_1-(z+z^{\star}+\lambda/2-\varepsilon)\right] =0,
\end{equation}
plus the normalization condition
\begin{equation}
\alpha^{\star}\alpha+3m_2=1.
\end{equation}

Manipulating these equations and reparameterizing them with
$w\equiv\varepsilon-z$, we get the following equivalent set of
equations:
\begin{eqnarray}
0&=&a(2w-\lambda)+b(2w^{\star}-\lambda)-m_1\varepsilon+n_1,
\label{a2w}\\
0&=&\frac{3\lambda}{2}\left[ a(w-\lambda/2)+
b(w^{\star}-\lambda/2)\right]
+4w^{\star}w(a+b-m_1/2)-\frac{3\lambda}{4}(w+w^{\star}
-\lambda/2)m_1,\\
0&=&w^{\star}+w-6(\varepsilon m_2-n_2)-3\lambda\left[ 2(a^{\star}
+b^{\star})m_1+2m_1^{\star}(a+b)-m_1^{\star}m_1\right] ,\\
0&=&\frac{1}{4}-4(a^{\star}+b^{\star})(a+b)-\frac{3m_2}{4}
-m_1^{\star}m_1+2\left[ (a^{\star}+b^{\star})m_1+
m_1^{\star}(a+b)\right], \label{1div4}\\
\alpha&=&2m_1-4(a+b),\\
\tau&=&(\varepsilon-w^{\star})(\varepsilon-w)-
\frac{3\lambda^2}{16}\, .
\end{eqnarray}

The first four equations of this set should determine the
parameters $a$, $b$, and $w$, while the last two equations
simply give us the values of $\alpha$ and $\tau$ in terms of
these parameters. To get closed relations between $a$, $b$,
and $w$, we have to substitute in Eqs.\
(\ref{a2w})--(\ref{1div4}) the explicit expressions of $m_1$,
$m_2$, $n_1$, and $n_2$ in terms of $a$, $b$, and $w$, namely,
\begin{eqnarray}
m_1&=&aM+bM^{\star},\\
n_1&=&a\left[ \lambda+(\varepsilon-w)M\right]
+b\left[\lambda+(\varepsilon-w^{\star})
M^{\star}\right] ,\\
m_2&=&a^{\star}bB^{\star}+b^{\star}aB+\lambda\frac{a^{\star}a
+b^{\star}b}{w^{\star}-w}(M-M^{\star}),\\
n_2&=&ab^{\star}\left[\lambda M+(\varepsilon-w)B\right]
+a^{\star}b\left[\lambda M^{\star}+(\varepsilon-
w^{\star})B^{\star}\right] +\nonumber\\
& &\lambda\frac{a^{\star}a+b^{\star}b}{w^{\star}-w}
\left[ (\varepsilon-w)M-(\varepsilon-w^{\star})M^{\star}\right] ,
\end{eqnarray}
where $M$ and $B$ are given by
\begin{equation}
M=\frac{\lambda}{\Gamma}\int\frac{d{\bf q}}{\varepsilon_{{\bf q}}
-\varepsilon+w}\; ,
\; B=\frac{\lambda^2}{\Gamma}\int\frac{d{\bf q}}
{(\varepsilon_{{\bf q}}-\varepsilon+w)^2}\, .
\end{equation}

The result of this substitution is, after some manipulations,
\begin{eqnarray}
0&=&{\rm Im}\left[w^{\star}(8w^{\star}w+3\lambda(1+2/A)
w^{\star}-3\lambda^2/2)\right],\label{8ww}\\
0&=&{\rm Im}\left[ w^{\star 2}A^{\star}[(4A+3B/A)(w^{\star}
-w)^2+3\lambda(A+2)(w^{\star}-3w)+6\lambda(1-A/A^{\star})w]
\right] , \\
a^{\star}a&=&\frac{{\rm Im}[w^{\star 2}w]}{{\rm Im}\left[
w^{\star}A^{\star}(8Aw^{\star}w-(4A+3B/A)w^{\star}(w^{\star}-w)
-6\lambda w)\right]}\, ,\\
b&=&-\frac{wA}{w^{\star}A^{\star}}a\, ,\label{bofa}
\end{eqnarray}
where $A\equiv M-2$.

It should be noted that Eqs.\ (\ref{8ww})--(\ref{bofa}) determine
everything except the phase of $a$. This is understandable
since a change of this phase leads to the multiplication of the
state by a global phase factor. Thus, we can fix the phase of $a$
by demanding, for example, that $b=a^{\star}$. This selection
yields a real wave function and gives rise to the final equations
listed in Section IV.

$A$ and $B$ are given, in terms of the density of states $D(\eta)$
of the normalized band $\eta_{{\bf k}}$, by
\begin{equation}
A=\lambda\int^{1/2}_{-1/2}\frac{D(\eta)\,d\eta}
{(1-|\lambda|)\eta-\varepsilon+w}-2\; ,\;
B=\lambda^2\int^{1/2}_{-1/2}\frac{D(\eta)\,d\eta}
{\left[ (1-|\lambda|)\eta-\varepsilon+w\right]^2}\, .
\end{equation}

For the model density selected in Section IV ($D(\eta)=
\frac{8}{\pi}\sqrt{1/4-\eta^2}\,$), these expressions can
be integrated analytically. We get, for ${\rm Im}[w]<0$,
\begin{eqnarray}
A&=&\frac{8\lambda}{(1-|\lambda|)^2}\left[w-\varepsilon+i
\sqrt{(1-|\lambda|)^2/4-(\varepsilon-w)^2}
\right]-2,\label{aint}\\
B&=&\frac{-8\lambda^2}{(1-|\lambda|)^2}
\left[1+\frac{i(\varepsilon-w)}{\sqrt{(1-|\lambda|)^2/4-
(\varepsilon-w)^2}}\right].\label{bint}
\end{eqnarray}

\vfill
\newpage
\begin{figure}
\caption{Zero-temperature spectrum of the
antiferromagnetic Kondo lattice with one conduction electron,
in the lowest approximation, for decreasing values of $\lambda$.
The dashed curve parameterizes the values of the bare conduction
band and the solid curves correspond to the quadruplet
(horizontal line) and the two doublets.
For low values of $\lambda$, the doublets give rise to a
band that closely follows the bare conduction states and a
practically flat, heavy-fermion band.}
\end{figure}

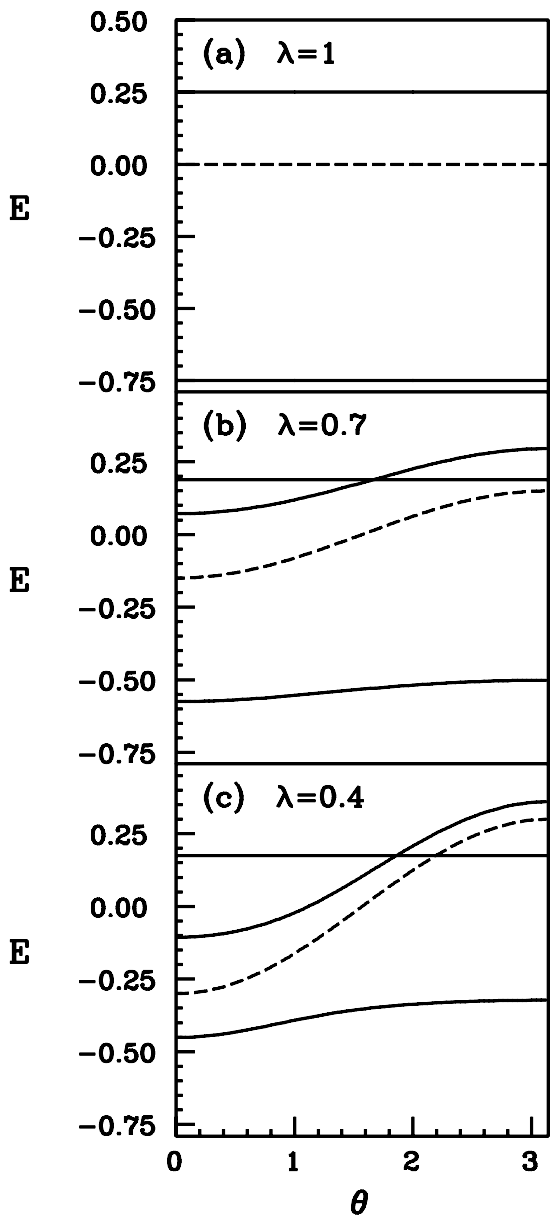
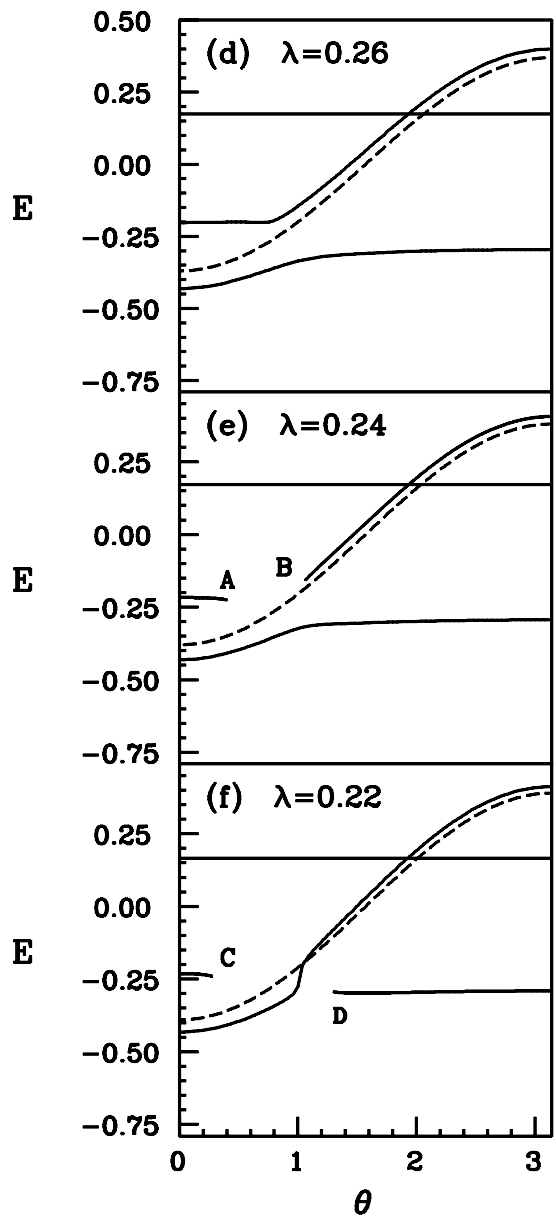
\end{document}